\documentclass[letterpaper, 10pt, conference]{IEEEtran}  % Comment this line out if you need a4paper

\IEEEoverridecommandlockouts                              % This command is only needed if 
\usepackage{setspace}
\usepackage{graphicx}
\usepackage[hyphens]{url}
\usepackage{algorithm}
\usepackage{algpseudocode}
\usepackage{amsfonts}
\usepackage{amsmath}
\usepackage{cite}
\usepackage{array}

\usepackage{float}
\usepackage{stfloats}
\usepackage{comment}
\usepackage{multirow}

\usepackage{bmpsize}

\usepackage{color,soul}
\usepackage{amssymb}

\title{Network--Co-Designed Dynamic Sparsification for Dependable Distributed Robot Learning over 3GPP Networks}

\author{
    \IEEEauthorblockN{G{\'e}za Szab{\'o}\IEEEauthorrefmark{1}, Boldizs{\'a}r Bunda\IEEEauthorrefmark{1}}
    \IEEEauthorblockA{\IEEEauthorrefmark{1}Ericsson Hungary Ltd., Budapest, Hungary, 
    email: \{geza.szabo\}@ericsson.com}
}

\usepackage[acronym, nomain]{glossaries}
\newacronym{3GPP}{3GPP}{Third Generation Partnership Project}
\newacronym{5G}{5G}{fifth generation of mobile communications}
\newacronym{5GC}{5GC}{5G Core}
\newacronym{5GS}{5GS}{5G System}
\newacronym{AAS}{AAS}{Asset Administration Shell}
\newacronym{AMF}{AMF}{Access and Mobility Management Function}
\newacronym{AES}{AES}{Application Enabler Server}
\newacronym{AF}{AF}{Application Function}
\newacronym{AGV}{AGV}{Automated Guided Vehicles}
\newacronym{AI}{AI}{Artificial Intelligence}
\newacronym{API}{API}{Application Programming Interface}
\newacronym{ARIAC}{ARIAC}{Agile Robotics for Industrial Automation Competition}
\newacronym{AS}{AS}{Application Server}
\newacronym{BS}{BS}{Base Station}
\newacronym{CapEx}{CapEx}{Capital Expenditure}
\newacronym{CAPIF}{CAPIF}{Common API Framework}
\newacronym{CPU}{CPU}{Central Processing Unit}
\newacronym{CPS}{CPS}{Cyber Physical System}
\newacronym{CPPS}{CPPS}{Cyber Physical Production System}
\newacronym{CSP}{CSP}{Communication Service Provider}
\newacronym{CSR}{CSR}{Customer Service Request}
\newacronym{DT}{DT}{Digital Twin}
\newacronym{EMD}{EMD}{Earth Mover's Distance}
\newacronym{EMOS}{EMOS}{Estimated Mean Opinion Score}
\newacronym{ERP}{ERP}{Enterprise Resource Planning}
\newacronym{FL}{FL}{Federated Learning}
\newacronym{GPU}{GPU}{Graphics Processing Unit}
\newacronym{GSMA}{GSMA}{Global System for Mobile Communications Association}
\newacronym{GUI}{GUI}{Graphical User Interface}
\newacronym{HTTP}{HTTP}{HyperText Transfer Protocol}
\newacronym{HW}{HW}{Hardware}
\newacronym{IAT}{IAT}{Inter-Arrival Time}
\newacronym{ID}{ID}{Identifier}
\newacronym{IEC}{IEC}{International Electrotechnical Commission}
\newacronym{IEEE}{IEEE}{Institute of Electrical and Electronics Engineers}
\newacronym{IIoT}{IIoT}{Industrial Internet of Things}
\newacronym{IoT}{IoT}{Internet of Things}
\newacronym{IT}{IT}{Information Technology}
\newacronym{IP}{IP}{Internet Protocol}
\newacronym{JSON}{JSON}{JavaScript Object Notation}
\newacronym{KPI}{KPI}{Key Performance Indicator}
\newacronym{LIDAR}{LIDAR}{Laser Imaging, Detection, And Ranging}
\newacronym{LoS}{LoS}{Line of Sight}
\newacronym{MAC}{MAC}{Medium Access Control}
\newacronym{MBB}{MBB}{Mobile Broadband}
\newacronym{MaaS}{MaaS}{Metal-as-a-Service}
\newacronym{MES}{MES}{Manufacturing Execution System}
\newacronym{MIMO}{MIMO}{Multiple-Input Multiple-Output}
\newacronym{ML}{ML}{Machine Learning}
\newacronym{MOS}{MOS}{Mean Opinion Score}
\newacronym{MQTT}{MQTT}{Message Queuing Telemetry Transport}
\newacronym{MTC}{MTC}{Machine Type Communication}
\newacronym{NEF}{NEF}{Network Exposure Function}
\newacronym{NF}{NF}{Network Function}
\newacronym{NFV}{NFV}{Network Function Virtualization}
\newacronym{NPN}{NPN}{Non-Public Network}
\newacronym{NRM}{NRM}{Network Resource Management}
\newacronym{NWDAF}{NWDAF}{Network Data Analytics Function}
\newacronym{OFDM}{OFDM}{Orthogonal Frequency Division Multiplexing}
\newacronym{OM}{OM}{Operations and Maintenance}
\newacronym{OPC}{OPC}{Open Platform Communications}
\newacronym{OpEx}{OpEx}{Operating Expenses}
\newacronym{OS}{OS}{Operating System}
\newacronym{OT}{OT}{Operational Technology}
\newacronym{ONVIF}{ONVIF}{Open Network Video Interface Forum}
\newacronym{PCF}{PCF}{Policy Control Function}
\newacronym{PLC}{PLC}{Programmable Logic Controller}
\newacronym{PPO}{PPO}{Proximal Policy Optimization}
\newacronym{PT}{PT}{Physical Twin}
\newacronym{PubSub}{PubSub}{Publish Subscribe}
\newacronym{QCI}{QCI}{QoS Class Identifiers}
\newacronym{QoE}{QoE}{Quality of Experience}
\newacronym{QoS}{QoS}{Quality of Service}
\newacronym{QoSensing}{QoSensing}{Quality of Sensing}
\newacronym{RAN}{RAN}{Radio Access Network}
\newacronym{RL}{RL}{Reinforcement Learning}
\newacronym{ROS}{ROS}{Robot Operating System}
\newacronym{RTDE}{RTDE}{Real-Time Data Exchange}
\newacronym{SBI}{SBI}{Service-Based Interface}
\newacronym{SBA}{SBA}{Service-Based Architecture}
\newacronym{SCEF}{SCEF}{Service Capability Exposure Function}
\newacronym{SDN}{SDN}{Software Defined Networking}
\newacronym{SDR}{SDR}{Software Defined Radio}
\newacronym{SEAL}{SEAL}{Service Enabler Architecture Layer}
\newacronym{SIM}{SIM}{Subscriber Identity Module}
\newacronym{SLAM}{SLAM}{Simultaneous Localization and Mapping}
\newacronym{SMF}{SMF}{Session Management Function}
\newacronym{SW}{SW}{Software}
\newacronym{TDMA}{TDMA}{Time-Division Multiple Access}
\newacronym{TCP}{TCP}{Transmission Control Protocol}
\newacronym{TS}{TS}{Technical Specifications}
\newacronym{TSG}{TSG}{Technical Specification Group}
\newacronym{TSN}{TSN}{Time-Sensitive Networking}
\newacronym{UA}{UA}{Unified Architecture}
\newacronym{UE}{UE}{User Equipment}
\newacronym{URLLC}{URLLC}{Ultra Reliable Low Latency Communications}
\newacronym{VAL}{VAL}{Vertical Application Layer}
\newacronym{VASEAL}{VASEAL}{Vertical Application Service Enabler Architecture Layer}
\newacronym{VM}{VM}{Virtual Machine}
\newacronym{WiFi}{WiFi}{Wireless Fidelity}
\newacronym{WG}{WG}{Working Group }

\begin{document}

%\date{}

\maketitle
%\begin{center}
%\LARGE Network--Co-Designed Dynamic Sparsification for Dependable Distributed Robot Learning over 3GPP Networks \\
%\large G{\'e}za Szab{\'o}, Boldizs{\'a}r Bunda, J{\'o}zsef Varga, Marcell Balogh \\
%\small Ericsson Hungary Ltd., Budapest, Hungary \\
%\end{center}
%\vspace{1ex}

\thispagestyle{empty}

% Dependable networks
% https://www.ericsson.com/en/reports-and-papers/ericsson-technology-review/articles/dependable-networks

%TBD: update
\begin{abstract}
Robot fleets performing distributed learning often face volatile uplink conditions like bandwidth variation, packet loss, and latency spikes that delay or destabilize training. We propose a network--co-designed dynamic sparsification loop that couples model-update compression to network-facing byte budgets and time constraints. We describe two realizations: (i) a standards-aligned SA6/5GC mapping that leverages CAPIF with NEF/PCF/NWDAF, enriched with AI-specific semantics; and (ii) a VASEAL-style role-reversed loop where the learning system exposes learning-state feedback and receives bounded, time-limited control intents. We implement the mechanism in an asynchronous federated reinforcement learning framework with heterogeneous simulated quadruped robots and reproducible uplink impairments via emulated network links. Results indicate that substantial communication reduction is achievable under dynamic constraints without materially degrading learning progress in the studied setting.
\end{abstract}

\IEEEpeerreviewmaketitle

\section{Introduction}

Robotics and \Gls{AI} are converging toward dependable industrial systems in which fleets of mobile robots must coordinate across dynamic factory, logistics, and field environments. End-to-end digital twins are increasingly used to train, validate, and stress-test such systems before deployment, reflecting a shift from isolated autonomy functions toward robust and trustworthy operation across the full pipeline.

A central challenge is scalability under heterogeneity. Robots face different terrain, sensor noise, dynamics, and failure modes, so classical centralized training is insufficient. Instead, distributed multi-agent learning allows each robot to improve from local experience while contributing to a shared model. This introduces system-level challenges as robots learn at different rates and communicate over uncertain or degraded wireless links.

In these systems, communication is a first-order factor: bandwidth drops, latency spikes, congestion, and intermittent connectivity directly affect convergence and predictability. At the same time, modern \Gls{3GPP} systems expose observability and control surfaces (\Gls{NEF}, \Gls{PCF}, \Gls{NWDAF}) that remain underused by \Gls{AI} applications. We therefore pursue a network--co-designed approach in which dynamic sparsification is coupled to network-exposed budgets, deadlines, and analytics, closing the loop between model-update generation and network conditions while remaining compatible with current \Gls{5GC} mechanisms.

Our contributions are threefold:
(i) we define a concrete closed-loop formulation for bandwidth-aware update generation, where clients adapt sparsification to satisfy explicit time-bounded byte budgets and guardrails derived from network conditions;
(ii) we provide two complementary realization paths for this loop: a standards-aligned 3GPP SA6 5GC mapping that enriches existing NEF/PCF/NWDAF exposure and control with AI-specific semantics, and a \Gls{VASEAL}-style~\cite{10164106} role-reversed realization where the learning system exposes learning-state feedback and receives bounded, time-limited control intents; and
(iii) we implement an end-to-end asynchronous federated reinforcement learning prototype with heterogeneous simulated quadruped robots, and evaluate the effect of adaptive sparsification under reproducible, emulated uplink variability, showing that substantial communication reduction can be achieved without degrading learning progress in the studied setting.

\section{Related work}\label{sec:related-work}

Recent work has combined federated or distributed \Gls{RL} with wireless and edge systems, but communication is usually treated as a coarse background constraint rather than as a controlled experimental variable. Among these studies, \cite{Xu2024_FedMultiAgentRL_GradientBounds} is the closest to our work: it studies federated independent-RL agents and introduces communication-efficient update exchange with convergence guarantees. However, it does not model volatile wireless behavior such as packet loss, delay variation, or connectivity heterogeneity, and it does not evaluate asynchronous robot-control clients. In contrast, our work focuses on end-to-end evaluation of asynchronous multi-robot learning under reproducible uplink variability, with adaptive compression tied directly to network conditions.

Our design also builds on standard 3GPP exposure and analytics capabilities. 3GPP~TS~23.501~\cite{3GPP_TS_23_501_2025} defines \Gls{NEF} and \Gls{PCF} for capability exposure and policy or \Gls{QoS} control, while 3GPP~TS~23.288~\cite{3GPP_TS_23_288_2025} defines \Gls{NWDAF} for network analytics. Industrial exposure proposals such as \Gls{SEAL} further show how these capabilities can be packaged for vertical applications \cite{3gppTS23434}. We therefore treat the network as an observable and controllable component of the learning system rather than as passive transport.

\section{AI-Oriented Exposure Proposals}
This section presents three complementary realizations of the closed-loop dynamic sparsification concept: a mapping onto existing 3GPP exposure capabilities, a set of AI-specific SA6-style abstractions, and a VASEAL-style role-reversed variant.

\subsection{Realization with existing 3GPP Exposure Capabilities}
The closed loop requires bidirectional exchange between the distributed learning application and the 3GPP network: the application exposes communication intent such as upload windows or QoS needs, while the network returns state and analytics such as throughput, latency, loss, or congestion indicators. In current 3GPP practice, this can be realized by an \Gls{AF} invoking CAPIF-discovered exposure services that map to NEF, PCF, and NWDAF functions, largely following the Network Resource Adaptation and Event Monitoring APIs of 3GPP~TS~23.434 \cite{3gppTS23434}.

This baseline is viable, but it lacks AI-specific semantics. The learning application must still translate network-facing capabilities into update budgets, pacing, and compression decisions. We therefore propose complementary AI-specific exposure abstractions tailored to distributed learning.

\subsection{Proposed AI-Oriented Exposure Procedures fitting in the 3GPP SA6 Framework}

We propose complementary AI-specific exposure abstractions, implementable as SA6 \emph{Application Enabler} APIs, to address recurring pain points in our target use case: heterogeneous quadruped robots learn at different speeds due to terrain and compute variability, while uplink conditions fluctuate due to congestion and radio dynamics.
In round-based (synchronous) FL implementations--still a common default in frameworks like Flower \cite{flower} and baselines--clients tend to upload in bursts and the system becomes straggler-limited, wasting airtime and delaying aggregation. In asynchronous operation, the main risks shift toward stale updates, unfairness (fast clients dominating), and uncontrolled uplink contention. In both cases, large model updates can exceed the instantaneous capacity, causing queueing, retries, and missed ``useful-by'' deadlines, which destabilizes learning progress. Additionally, learning traffic is multi-class (model updates vs.\ control/telemetry/metrics) but is often indistinguishable to the network without explicit semantics, leading to suboptimal QoS treatment. Finally, neither side has a clean contract: the application guesses what the network can deliver, while the network lacks visibility into learning phase/criticality and admissible compression ranges. %The following APIs provide standardized objects for time-bounded byte budgets, traffic marking, learning-phase signaling, analytics feedback, cohort pacing, and budget-to-compression guidance to close this gap.
We therefore propose a complementary set of SA6 Application Enabler APIs (illustrative names), exposed via CAPIF by an Application Enabler Server and consumed by the learning AF:
\begin{itemize}
    \item AI Update-Window Scheduling (AI-UWS): the learning AF requests per-UE, time-bounded upload windows with explicit byte budgets and deadlines (e.g., \texttt{CreateWindowRequest}, \texttt{CancelWindow}, and a \texttt{WindowDecision} callback). The network admits the window, provides a feasible alternative, or rejects based on load.
    \item AI Gradient-Flow Marking (AI-GFM): the AF registers learning-related traffic descriptors (gradient/model-update/control/telemetry) for consistent classification (e.g., \texttt{RegisterFlow}, \texttt{UpdateFlow}, \texttt{DeleteFlow}), enabling per-flow QoS treatment (QFI/5QI) without embedding network-specific selectors into the learning code.
    \item AI Model-Change Notification (AI-MCN): the AF signals learning phase and criticality changes (e.g., \texttt{NotifyModelPhase} with critical round, evaluation, recovery), allowing time-limited policy adjustments (temporary uplift or deferral) to stabilize progress under volatile connectivity.
    \item AI Resiliency Analytics Contract (AI-RAC): the AF subscribes to a learning-relevant KPI bundle (e.g., \texttt{(Un)/SubscribeAnalytics}, \texttt{AnalyticsReport} callback) such as throughput percentiles, latency/jitter, loss, upload success ratio, and ``budget adherence'' for scheduled windows.
    \item AI Cohort Coordination (AI-CC): the AF requests group-level pacing and staggering policies (e.g., \texttt{CreateCohort}, \texttt{RequestStaggerPlan}, \texttt{StaggerPlanReport}) to cap concurrent uploads and reduce burstiness while preserving fairness across heterogeneous robots.
    \item AI Budget-to-Compression Guidance (AI-BCG): a domain-specific API for translating network-facing budgets into learning-facing constraints (e.g., \texttt{GetBudgetGuidance} or \texttt{BudgetGuidanceReport}), returning recommended byte budgets and admissible compression ranges (keep fraction, quantization bounds, delta mode) consistent with current forecasts, reducing per-client heuristics and improving interoperability.
\end{itemize}

In summary, all proposed AI-oriented SA6 APIs can be viewed as thin, domain-specific wrappers (carrying learning semantics such as byte budgets, phases, and admissible compression ranges) around existing 5GC exposure and control capabilities. %CAPIF provides discovery, authentication/authorization, and onboarding of the exposure services; NEF provides northbound invocation and eventing; PCF realizes short-lived QoS/policy adjustments for the registered learning flows (e.g., update windows, prioritization, and pacing constraints); and NWDAF provides the predictive analytics (throughput/load/experience) that drive closed-loop adaptation. 
Among the proposals, AI-UWS, AI-GFM, AI-MCN, AI-CC primarily map to policy and traffic treatment (NEF/PCF, plus optional scheduling at the AF), whereas AI-RAC and AI-BCG are analytics-driven: they consume NWDAF/NEF insights and translate them into application-level feedback \Glspl{KPI} and compression guardrails.

While the above presents an SA6-style, network-provided enabler view -- CAPIF-exposed APIs consumed by the learning AF --, the same closed-loop objects can also be realized in a VASEAL-style role-reversed manner.

\subsection{Proposed AI-Oriented Exposure Procedures fitting in the VASEAL Framework}\label{sec:proposed-system}

\begin{figure*}[tb]
\centering
\includegraphics[width=\textwidth]{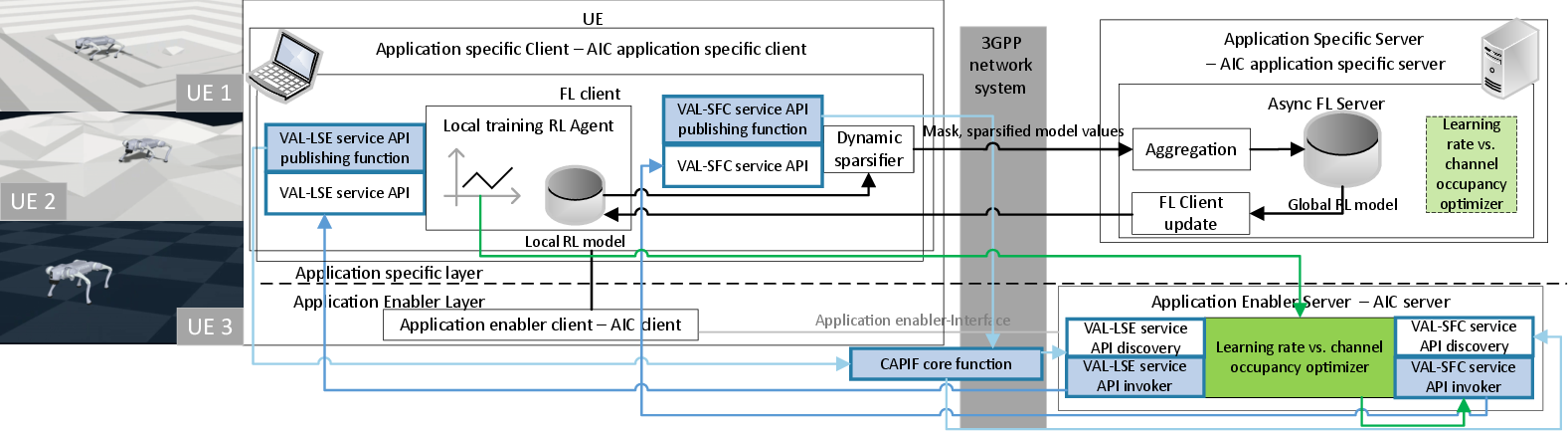}
\caption{The proposed closed-loop architecture called AI-Control (AIC) fitted in 3GPP SA6 Application Enabler Layer and aligned in the network as the application exposure consumer concept}
\label{fig:system}
\end{figure*}

In the VASEAL architecture~\cite{10164106}, the controller functionality is realized by a \Gls{NF}, while the exposure provider is an \Gls{AF}. In our setting the \Gls{FL} system (\Gls{VAL}/AF) acts as the API provider by exposing learning-state signals (e.g., a convergence/learning-rate proxy, update staleness, and upload success), while a network-side function or an operator application consumes these signals and issues bounded, time-limited control intents e.g.,~sparsification/filtering targets. 
The exposed objects remain similar to the SA6-style proposals (byte budgets, admissible compression ranges, phase/criticality events, and KPIs), but the direction of consumption is inverted: the network uses learning feedback to drive QoS/policy decisions, and the learning application returns feasibility and safety constraints to prevent harmful control during fragile training phases.

We therefore propose a complementary set of CAPIF-exposed VAL-provided APIs:
\begin{itemize}
    \item VAL Learning-State Exposure (VAL-LSE): network subscribes to learning KPIs and events (e.g., \texttt{SubscribeLearningMetrics} and \texttt{LearningMetricsReport}) such as learning rate/convergence proxy, update staleness, and upload success ratio.
    \item VAL Sparsification and Filtering Control (VAL-SFC): network issues a time-bounded control intent (e.g., \texttt{SetFilteringRatio} or \texttt{SetCompressionPolicy}) with guardrails (min keep fraction, max quantization) and receives an acknowledgement or feasibility response.
    %\item \textbf{VAL Feasibility and Guardrails (VAL-FG):} the learning system returns admissible ranges and recommended settings (e.g., \texttt{GetAdmissibleCompressionRange}), enabling the network to avoid harmful policies during fragile training phases.
\end{itemize}

\section{Proposed procedure}
%We use the end-to-end distributed quadruped-learning setup described in Section~\ref{sec:implementation} as the running example to illustrate how the proposed exposure procedures support dependable learning under heterogeneous compute/terrain and volatile uplink conditions.

%The procedure emphasizes the \emph{closed loop} rather than the platform details. 
The learning system exposes lightweight learning-state feedback (e.g., learning-rate/convergence proxy, update staleness, upload success) via VAL-LSE, while a network-side controller combines this feedback with available network analytics (when deployed) to produce bounded, time-limited control intents via VAL-SFC (e.g., filtering/keep fraction, quantization bounds, optional byte-budget guidance). Clients then enforce the intent locally by adjusting sparsification and compression before each upload. The resulting feasibility acknowledgement and outcome metrics feed the next control step. %In our experiments, network variability is emulated using Linux traffic control, but the same control objects map to NEF/PCF/NWDAF-backed actuation in a live 5GC deployment.

\begin{figure}[tb]
\centering
\includegraphics[width=\columnwidth]{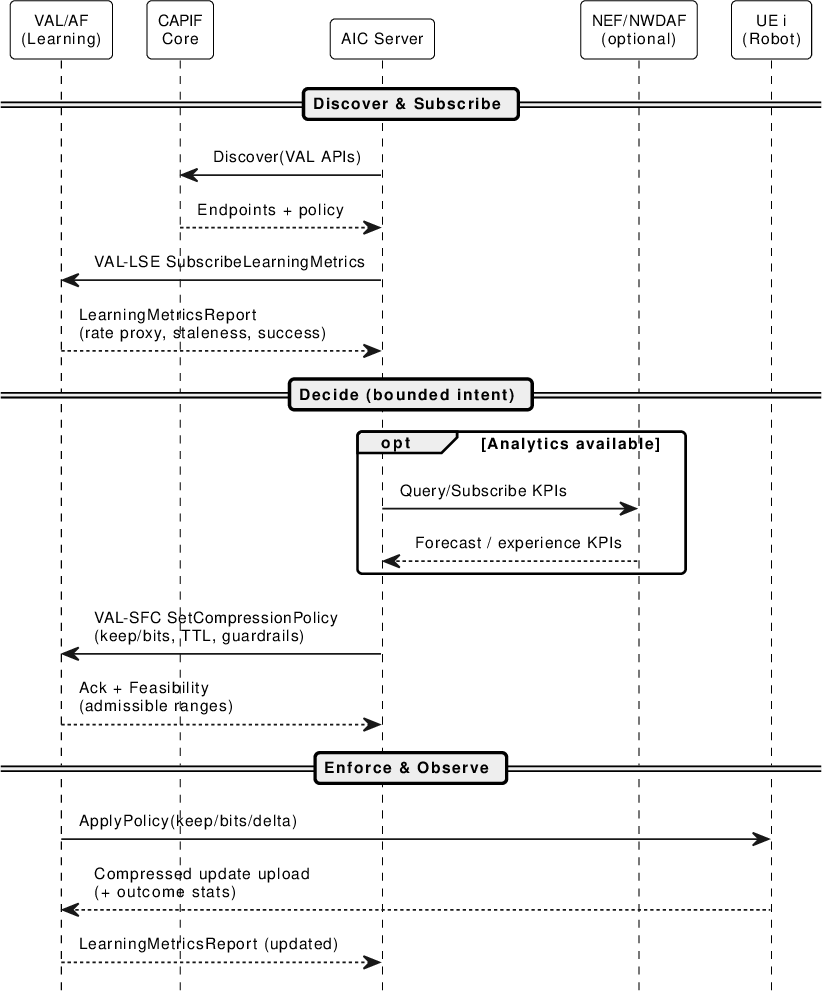}
\caption{The sequence diagram of the proposed closed-loop AI-Control (AIC) procedure fitted in VASEAL framework}
\label{fig:VASEAL_seq}
\end{figure}

The architecture is illustrated in Figure~\ref{fig:system}, showing how the VAL-AI Control (AIC) functions integrate into the VASEAL framework. Figure~\ref{fig:VASEAL_seq} complements this view by summarizing the corresponding CAPIF-mediated message exchange between the learning system and SA6 application-enabler components.

\subsection{System Model}
We consider $N$ robots \Glspl{UE} that perform local \Gls{RL} and periodically transmit model updates to a central server over a 3GPP uplink. Uplink capacity varies over time due to radio conditions, mobility, and cell load, and the server aggregates updates asynchronously as they arrive (i.e., without a round barrier). For robot $i$ at time $t$, let $B_i(t)$ denote the predicted (or exposed) uplink bitrate and let $T_u(t)$ denote the duration of a short, scheduled upload opportunity (window). The corresponding effective byte budget is
\[
b_i(t) \triangleq \zeta\,B_i(t)\,T_u(t),
\]
where $\zeta\in(0,1)$ is a safety margin that accounts for protocol overhead and prediction error. Before each upload, the learning system selects compression parameters $\theta_i(t)$ (e.g., keep fraction, quantization bits, delta mode), which induce an encoded payload size $s_i(\theta_i(t))$. The key feasibility requirement is that the transmitted update fits within the available budget while retaining sufficient utility for learning progress.
In our prototype, $B_i(t)$ is approximated by a client-side estimate $\widehat{B}_i(t)$ obtained from the configured shaping limit and/or in-container measurements.

\subsection{Network--Co-Designed Dynamic Sparsification}
Given a dense-update size $S_{\text{full}}$ (bytes) and an upload budget $b_i(t)$, we select compression parameters by solving the constrained choice
\[
\begin{aligned}
    \theta_i^\star(t) &\in \arg\max_{\theta} \; U_i(\theta) \\
    \text{s.t.}\quad & s_i(\theta)\le b_i(t),
\end{aligned}
\]
where $U_i(\theta)$ is an application-level utility proxy (e.g., preserving the largest-magnitude parameter changes) and $s_i(\theta)$ is the corresponding encoded size. In practice we use
$\theta_i(t)=\{\rho_i(t),\,q_i(t),\,\text{delta}\}$, where $\rho_i(t)$ is the keep fraction, $q_i(t)$ the quantization bit-width (bounded by $b_{\max}$), and \texttt{delta} indicates whether delta encoding is enabled.

As a simple budget-driven heuristic, we set the keep fraction proportional to the budget relative to a dense update,
\[
\rho_i(t)=\min\!\left(1,\frac{b_i(t)}{S_{\text{full}}}\right),
\]
and reduce the quantization bit-width as sparsification becomes more aggressive:
\[
q_i(t)=\mathrm{clip}\!\Bigl(b_{\max}-\kappa\cdot\bigl(1-\rho_i(t)\bigr)\Bigr),
\]
where $\kappa$ controls the bits--sparsity trade-off and $\mathrm{clip}(\cdot)$ enforces valid bounds. Delta encoding is enabled for small budgets to further reduce payload size.

\section{Implementation in Our E2E Framework}\label{sec:implementation}

We implement the proposed closed-loop learning-over-network procedure (Section~\ref{sec:proposed-system} and Section~\ref{sec:implementation}) in a containerized end-to-end framework. Each robot learner runs in an isolated GPU/CPU-backed container with a terrain-specific configuration, while a central orchestration server performs asynchronous aggregation of whichever client updates arrive (no round barrier). This setup is used to evaluate how adaptive sparsification behaves under controlled uplink variability.

The simulation environment is built on top of Genesis~\cite{Genesis2023}, an open-source high-fidelity robotics simulator. Each robot executes a local \Gls{PPO}~\cite{schulman2017} training loop that collects experience, computes policy updates, and prepares an upload payload (dense or sparsified) according to the current control intent / budget.

The \Gls{FL} orchestration and aggregation are implemented using Flower~\cite{flower}. We extend Flower with a custom asynchronous server loop and strategy components to (i) accept client results opportunistically, (ii) decode sparse payloads when enabled, and (iii) update the global model immediately upon reception.

To emulate non-ideal 3GPP uplink conditions in a reproducible way, we inject network impairments using Linux traffic control (\texttt{tc})~\cite{tc}. Per-container shaping profiles constrain the uplink to represent congestion and time-varying capacity. 
These impairment profiles define the experimental conditions in Section~\ref{sec:evaluation}, where we study the stability and learning impact of bandwidth-aware sparsification. In a production deployment, the same control loop would consume NEF/NWDAF analytics and realize time-bounded QoS intents via PCF, but here we focus on an end-to-end, controllable test harness.

The VAL-AIC functions are integrated into the VASEAL framework described in Section~\ref{sec:proposed-system}. See~\cite{SZABO2024110581} for architectural details and the code in~\cite{opcua_code}.

\subsection{Reinforcement Learning's Reward Function}\label{sec:reward_function}

Each robot learns a locomotion policy via \Gls{PPO} using a composite reward that trades off task performance, stability, and control smoothness. At each control step ($\Delta t = 0.02\,\text{s}$), the reward includes: (i) a command-tracking term that encourages the base to follow the commanded planar linear velocity and yaw rate; and (ii) a set of regularizers that discourage undesired behavior, including excessive vertical oscillations (``bouncing''), deviation of the base height from a nominal clearance, abrupt action changes between consecutive steps, and large deviations from a default standing posture. The overall reward is scaled by $\Delta t$ so that cumulative return remains comparable across control frequencies.

This reward design promotes stable, energy-aware locomotion while remaining simple and robust across terrain-dependent difficulty. We use the same reward structure for all agents. Differences in achieved mean reward primarily reflect differences in terrain profiles rather than changes in the learning objective.

%In a production network, \texttt{tc}-based shaping is replaced by NEF/PCF/NWDAF integration: the same policy loop consumes network analytics and allocates upload windows via QoS flows, with the sparsifier enforcing the negotiated envelope.

%\subsection{Algorithm (Client--Server)}
%\begin{comment}
\begin{algorithm}[t]
\caption{Asynchronous Federated RL with Bandwidth-Aware Delta Sparsification}
\label{alg:grfl}
\begingroup
\footnotesize
\begin{algorithmic}[1]
\State \textbf{Server init:} load initial policy $\theta$; set aggregation weight $w$.
\State \textbf{Loop (server):} dispatch client training jobs concurrently; on any completion:
\State \hspace{1em} decode client result to obtain $\theta^{(c)}$ (dense or reconstructed)
\State \hspace{1em} update $\theta \leftarrow \theta + w(\theta^{(c)}-\theta)$ \hfill (Eq.~\ref{eq:async-update})
\Statex
\State \textbf{Client:} receive $\theta$; run one local RL interval to obtain $\theta_{\text{local}}$.
\If{sparse enabled}
  \State estimate $\widehat{B}_{\text{tx}}$; set keep fraction $\rho$ \hfill (Eq.~\ref{eq:keep-fraction})
  \State $\Delta \leftarrow \theta_{\text{local}}-\theta$
  \State choose threshold $\tau$ so that $\approx\rho$ fraction of coordinates satisfy $|\Delta|\ge\tau$
  \State create payload $\mathcal{P}$: mask $m=\mathbb{I}(|\Delta|\ge\tau)$ and values $\theta_{\text{local}}[m]$
  \State transmit $\mathcal{P}$
\Else
  \State transmit dense $\theta_{\text{local}}$
\EndIf
\end{algorithmic}
\endgroup
\end{algorithm}
%\end{comment}

\begin{figure}[t]
\centering
\includegraphics[width=\columnwidth]{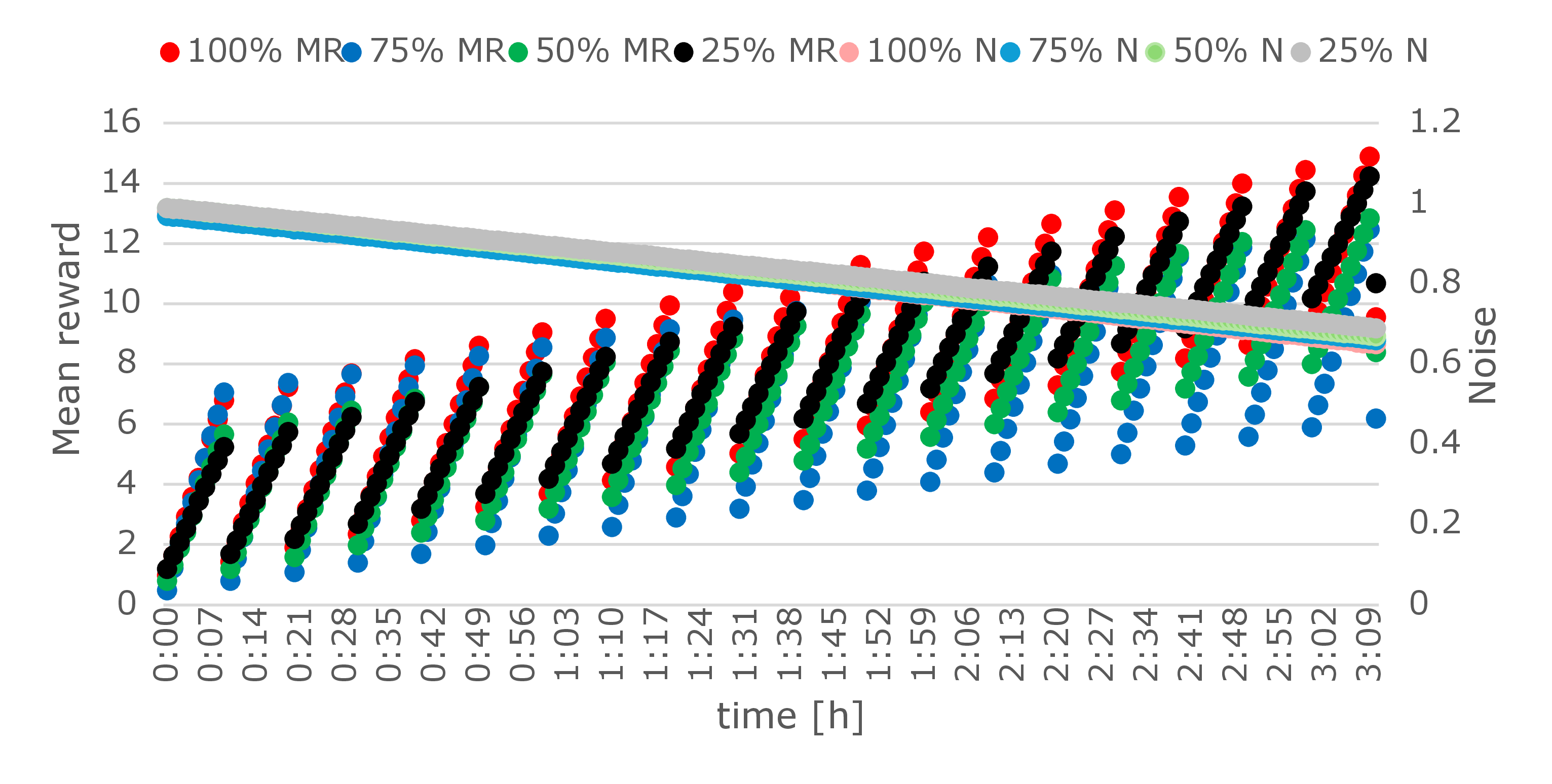}
%\caption{The mean reward (MR) and noise (N) across 4 heterogeneous robot learners under dynamic uplink constraints with various keep fractions for network--co-designed adaptive sparsification}
\caption{Mean reward and noise for four robots under different keep fractions}
\label{fig:example-run}
\end{figure}

\subsection{System Architecture}
We implement the system on top of Flower~\cite{flower} using an asynchronous client--server topology (Algorithm~\ref{alg:grfl}). A central server runs a custom asynchronous orchestration loop and a custom aggregation strategy to process client results as they arrive (no round barrier). Each client executes a local \Gls{RL} training routine and periodically uploads either (i) dense model parameters or (ii) a sparse, bandwidth-adaptive payload representing a partial update.

\paragraph{Server bootstrap}
At startup, the server initializes the global policy from an RL trainer checkpoint, converts the policy parameters into Flower-compatible parameters, and instantiates the selected asynchronous aggregation mode (e.g., step-size configuration and weighting rule).

\paragraph{Client loop}
Each client receives the current global parameters, performs one local training interval, and returns the resulting update in dense form or as a sparse payload (bitmask + selected values), depending on whether bandwidth-aware sparsification is enabled.

\subsection{Asynchronous Orchestration}
The server uses an event-driven asynchronous loop. Let $M$ be the number of connected clients.
The server dispatches \texttt{fit} jobs concurrently (thread pool) and updates the global model
immediately upon receiving any completed client result (no round barrier). This design naturally
handles heterogeneous client runtimes and stragglers.

\subsection{Asynchronous Aggregation Rule}
Let $\theta \in \mathbb{R}^d$ denote the current global parameters, and let $\theta^{(c)}$ be the
(reconstructed) client-side model produced after local training. Upon receiving an update, the
server applies a first-order asynchronous update of the form
\begin{equation}
\theta \leftarrow \theta + w\left(\theta^{(c)} - \theta\right),
\label{eq:async-update}
\end{equation}
where $w\in(0,1]$ is a mixing weight that controls how aggressively the global model moves toward the newly received client model. In the \emph{unweighted} (constant-weight) setting, we use a fixed $w$ for all clients and update arrivals (e.g., $w=0.5$), independent of client identity, local data volume, or staleness.

If multiple client results are processed together, the implementation first averages them element-wise to obtain
$\bar{\theta}$ and then substitutes $\theta^{(c)} \leftarrow \bar{\theta}$ in Eq.~\eqref{eq:async-update}.

\begin{figure}[t]
\centering
\includegraphics[width=.42\textwidth]{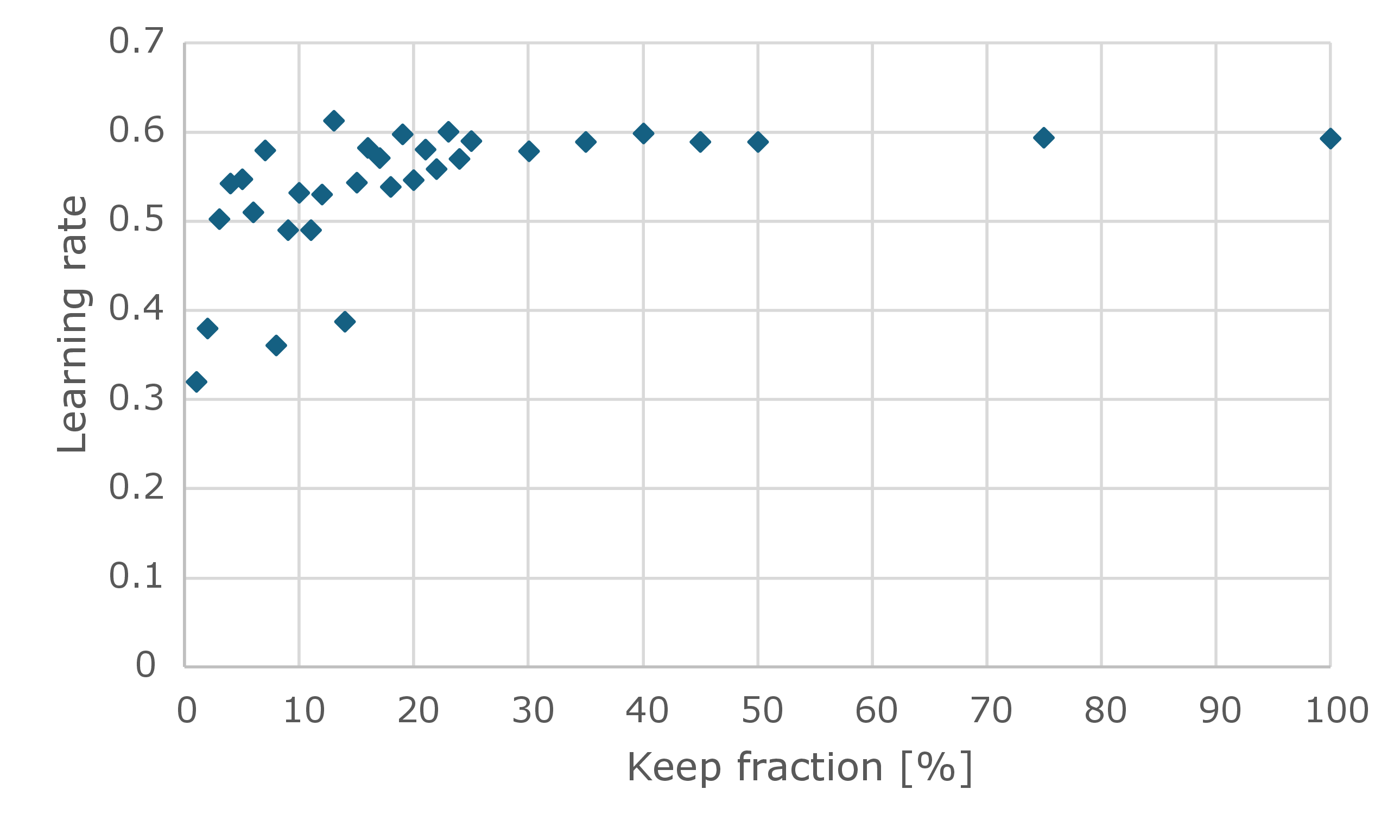}
%\caption{The impact of different minimum keep fractions on learning rate under dynamic uplink constraints with network--co-designed adaptive sparsification}
\caption{Effect of minimum keep fraction on learning rate under dynamic uplink constraints}
\label{fig:keepFractionVSlearningRate}
\end{figure}

\subsection{Bandwidth-Aware Delta Sparsification}
To reduce communication overhead, clients may transmit a sparse payload instead of full dense parameters. Sparsification is computed from parameter deltas $\Delta=\theta_{\text{local}}-\theta_{\text{global}}$, and adapts the transmitted fraction to the estimated uplink rate.

\paragraph{Bandwidth-to-keep mapping}
Clients estimate the available transmit rate from shaping configuration and/or in-container measurements. Let $\widehat{B}_{\text{tx}}$ denote the estimated uplink rate (bps). We map this estimate to a keep fraction via a bounded linear rule
\begin{equation}
\rho \;=\; \mathrm{clip}\!\left(\frac{\widehat{B}_{\text{tx}}}{B_{\max}},\; \rho_{\min},\; \rho_{\max}\right),
\label{eq:keep-fraction}
\end{equation}
where $\mathrm{clip}(x,a,b)=\min(\max(x,a),b)$, $B_{\max}$ is a reference rate, and $(\rho_{\min},\rho_{\max})$ bound the minimum and maximum transmitted fractions. Optionally, an exponential moving average is applied to smooth $\widehat{B}_{\text{tx}}$. In our prototype we approximate the windowed byte budget $b_i(t)=\zeta B_i(t)T_u(t)$ using a fixed upload-window duration and a client-side bitrate estimate, hence we express $\rho$ directly as a function of $\widehat{B}_{\text{tx}}$ in Eq.~\eqref{eq:keep-fraction}.

\paragraph{Global Top-$K$ selection by $|\Delta|$}
Across all eligible parameter tensors, we compute a single global threshold $\tau$ such that approximately a fraction $\rho$ of coordinates satisfy $|\Delta_i|\ge \tau$. Equivalently, if $N$ is the number of eligible coordinates and $K=\lceil \rho N\rceil$, then $\tau$ is the $(N-K+1)$-th order statistic of $\{|\Delta_i|\}_{i=1}^N$.

\paragraph{Payload semantics (partial overwrite)}
The support (mask) is selected by large delta magnitude ($|\Delta|$), but the transmitted values are the post-training weights $\theta_{\text{local}}$ at the selected coordinates (not the deltas). The payload therefore represents a sparse \emph{partial overwrite} of the base model.

\subsection{Sparse Encoding and Reconstruction}
For each tensor key $k$, the sparse payload contains: (i) a packed bitmask $m_k$ indicating selected coordinates, (ii) the corresponding values $v_k$, and (iii) minimal metadata (shape and dtype) needed for reconstruction. On the server side, reconstruction unpacks the bitmask and overwrites the selected coordinates:
\begin{equation}
\theta_k \leftarrow \mathrm{Overwrite}(\theta_k,\, m_k,\, v_k),
\end{equation}
where $\mathrm{Overwrite}(\cdot)$ replaces the entries of $\theta_k$ indicated by $m_k$ with the provided values $v_k$. The reconstructed dense model $\theta^{(c)}$ is then used in Eq.~\eqref{eq:async-update}.

\section{Evaluation}\label{sec:evaluation}
We evaluate the proposed network--co-designed adaptive sparsification using the end-to-end framework described in Section~\ref{sec:implementation} (four heterogeneous quadruped learners with asynchronous aggregation). For reproducibility, we emulate time-varying uplink constraints via Linux traffic control (\texttt{tc})~\cite{tc} and sweep the enforced minimum keep fraction ($\rho_{\min}\in\{1.0,0.75,0.5,0.25\}$). Each run lasts for a fixed duration. %we report mean reward (MR) and noise (N) per learner. %and summarize learning speed using the normalized area under the smoothed reward envelope (Fig.~\ref{fig:keepFractionVSlearningRate}).

Figure~\ref{fig:example-run} shows the time evolution of the mean reward (MR) and noise (N) for four different keep fraction ratios (100\%, 75\%, 50\%, and 25\%) during \Gls{FL}. MR values exhibit repeated growth phases separated by periodic drops, while the corresponding N values decrease gradually throughout the experiment. For each keep fraction, both metrics are reported at one-minute resolution, allowing a direct comparison of learning dynamics under different proportions of transmitted model parameters.

Note that the mean rewards differ across agents because their terrain profiles are heterogeneous, resulting in varying task difficulty and learning progress. An agent operating on more challenging terrain may achieve lower rewards than one on an easier surface, due to the reward definition for vertical torso motion (see Section~\ref{sec:reward_function}).

To quantify learning speed, we compute the area under the smoothed reward envelope, obtained via a moving-average filter to suppress episodic oscillations. The resulting metric is normalized by training duration and asymptotic performance, yielding a dimensionless score that captures both convergence speed and final reward level. The results are summarized in Figure~\ref{fig:keepFractionVSlearningRate}.

Interestingly, reducing the fraction of transmitted model parameters even down to 5\% does not significantly impact learning speed. This can be explained by the fact that model updates are highly skewed, with the majority of learning progress driven by a small subset of parameters exhibiting large changes. By transmitting only the most significant updates, the proposed scheme preserves the dominant learning signal while implicitly filtering low-magnitude, potentially noisy updates, resulting in comparable convergence behavior.

%\section{Discussion and Limitations}
%Our approach assumes the AF can obtain trustworthy bitrate predictions and that short QoS windows can be honored; extreme congestion may still force aggressive sparsification that harms accuracy. Mapping byte budgets to compression parameters is model-dependent; safeguards (minimum keep fraction, bounded quantization) are necessary. Finally, coordination across administrative domains (e.g., public networks) requires standardization of AI-oriented exposure profiles.

\section{Conclusions and Further Work}\label{sec:conclusions}
This paper presented a network--co-designed dynamic sparsification approach for dependable distributed robot learning over 3GPP systems. The core idea is to couple learning-side compression decisions to network-facing budgets and time constraints, so that updates are transmitted in a size- and time-aware manner under variable uplink conditions. We mapped this closed loop to two complementary realizations: (i) a standards-aligned SA6 view that can be implemented using existing 5GC exposure, policy, and analytics functions (NEF/PCF/NWDAF) enriched with AI-specific semantics; and (ii) a VASEAL-style role-reversed realization in which the learning application exposes learning-state feedback and receives bounded, time-limited control intents. We implemented the mechanism in an end-to-end asynchronous federated reinforcement learning framework with heterogeneous quadruped learners and reproducible network impairment injection, and used it to assess how bandwidth-aware sparsification affects training dynamics. The results indicate that substantial reduction of transmitted parameters can be achieved without degrading learning progress in the studied setting, supporting the practicality of the proposed closed-loop design.
As a further work the proposed procedures could be standardized.

%Future work will (a) integrate the loop with live NEF/PCF/NWDAF control in a 5GC testbed, (b) quantify the impact of mobility, handovers, and multi-cell load on both communication reliability and learning stability, and (c) formalize safety-aware guardrails (e.g., minimum keep fraction and bounded quantization) that adapt to learning phase and update staleness. Additional work is also needed to standardize the proposed AI-oriented exposure objects (byte budgets, phases/criticality, cohort pacing, and resiliency KPIs) and to evaluate their interoperability across vendors and deployments.
% ...existing code...
\bibliographystyle{./bibliography/IEEEtran}
\bibliography{./bibliography/IEEEabrv,./bibliography/IEEEexample}

\end{document}